\documentstyle[12pt]{article}
\begin{document}
\setlength{\baselineskip}{0.30in}

\newcommand{\beq}{\begin{equation}}
\newcommand{\eeq}{\end{equation}}
\newcommand{\beqa}{\begin{eqnarray}}
\newcommand{\eeqa}{\end{eqnarray}}
\newcommand{\lsim}{\begin{array}{c}\,\sim\vspace{-21pt}\\<\end{array}}
\newcommand{\gsim}{\begin{array}{c}\sim\vspace{-21pt}\\>\end{array}}

 {\hbox to\hsize{Feb. 1993 \hfill UM-93-02}}
 {\hbox to\hsize{ \hfill JHU-TIPAC-930001}}

\begin{center}
\vglue .06in
{\Large \bf Decoupling in Non-Perturbative }\\
{\Large \bf Background Fields: the Thermal Sphaleron}
\\[.5in]

\begin{tabular}{cc}
\begin{tabular}{c}
{\bf Thomas M. Gould\footnotemark[1]}\\[.05in]
{\it Department of Physics}\\
{\it The Johns Hopkins University}\\
{\it Baltimore MD 21218 }\\[.15in]
\end{tabular}
&
\begin{tabular}{c}
{\bf I.Z. Rothstein\footnotemark[2]}\\[.05in]
{\it The Randall Laboratory of Physics}\\
{\it University of Michigan}\\
{\it Ann Arbor MI 48109 }\\[.15in]
\end{tabular}
\end{tabular}

\footnotetext[1]{gould@dirac.pha.jhu.edu}
\footnotetext[2]{irothstein@umiphys.bitnet}
{Abstract}\\[-.1in]
\end{center}

\begin{quotation}
Standard decoupling of heavy fermions may fail when
there are non-perturbative variations in a scalar field which
gives masses to the fermions. One situation of phenomenological
relevance is the case of sphalerons in the presence of
fermions at finite temperatures. The free energy of a simple model
is determined using a non-perturbative technique to study
the effect of fermions on the scalar field. The effects
 of quantum and thermal fermionic fluctuations on the free energy
 of the thermal sphaleron are calculated, including
contributions from the gradients of the scalar field to all orders.

\end{quotation}

\newpage
\section{Introduction}

When a particle is made heavy by increasing a dimensionful
parameter,
its virtual effects on physical quantities are suppressed by
inverse powers of the parameter.
Under these conditions the particle is said to decouple\cite{APP}.
However, decoupling fails when a particle is made heavy by
increasing a dimensionless coupling.
This is the case for instance when a fermion mass is made heavy
by increasing a Yukawa coupling, while the Higgs vacuum expectation
value is held fixed.
Therefore,
chiral gauge theories in particular can be sensitive to the short
distance physics associated with heavy fermions,
as a consequence of this {\it perturbative} \mbox{non-decoupling\cite{DHO}.}

Recently,
Banks and Dabholkar have pointed out that decoupling may also fail when
there are non-perturbative variations in a scalar field which gives
masses to fermions\cite{BAN}.
More specifically,
these authors noted that the effects of heavy fermions cannot be
summarized by a local effective Lagrangian for the light degrees
of freedom if the scalar field which gives the fermion its mass
goes through a zero.
Heuristically,
this {\it non-perturbative} non-decoupling can be understood as stemming
from the local vanishing of the fermion mass.
To see this more clearly,
consider the fermion determinant in a scalar background $\phi(x)$.
The effective action for the the scalar field may be expressed as
\beq
\label{eq:intro}
S_{eff} \left[\phi\right] \: = \:
S\left[\phi\right] \: - \: i\,\ln \, det
\left[\, i\not\!\! D \, - \, g\,\phi \,\right] \, ,
\eeq
where $S\left[\phi\right]$ is the classical scalar field action,
$i\!\not\!\! D $ is a gauge covariant derivative
and $g$ is a Yukawa coupling.
To evaluate the determinant in an arbitrary background
one may attempt a gradient expansion. However, any such approximation is
surely doomed. The leading order term in such an expansion is given by
 $\left(\partial_\mu\phi\right)^2/g^2\phi^2$.
It is clear that this approach fails in the case of
a vanishing scalar field.
Thus,
the effects of a heavy fermion can only be summarized
by a {\it non-local} effective lagrangian.

This non-decoupling can be recognized by an apparent paradox\cite{BAN}.
Begin by considering a chiral gauge theory with heavy mirror fermions
for all light fermions.
The fermion number currents for the light fermions and for the mirror
fermions are assumed to be anomalous,
while their difference is exactly conserved.
Furthermore,
assume that at energies small compared to the heavy fermion mass,
there exists a local effective lagrangian which describes the physics
of the light degrees of freedom.
Now consider the decay of a light fermion through the fermion number anomaly.
In the constrained instanton background \cite{AFF},
the scalar will necessarily pass through zero, and
furthermore, {\it all} of the flavor eigenstates will have zero mode
solutions to the Dirac operator.
In the absence of mixing between heavy and light fermions,
a non-vanishing decay amplitude necessarily requires heavy fermions
on external legs to soak up their zero modes.
But, this contradicts the initial assumption  that a local effective theory
incorporates all of the physics of the heavy fermions.

This paper will consider the effects of heavy fermions on background scalar
fields with non-perturbative variations.
Specifically,
the effect of heavy fermions on sphalerons at finite temperature will be
addressed\cite{KLI}.

It is by now widely accepted that fermion number violation
occurs at finite temperatures due to the production and decay of
sphalerons\cite{KUZ,ARN}.
Sphaleron processes may then play an instrumental role in
determining the baryon asymmetry in the universe\cite{CKN}.
Considering the discussion given above,
it is not hard to imagine that heavy fermions may influence
the production and decay of sphalerons.
In the process of sphaleron production and decay,
the fermion number of each flavor of fermion must be violated,
as a consequence of the fermion number anomaly discussed in the case
of instantons above.
These fermions may receive masses exceeding the mass of the
classical sphaleron through Yukawa couplings to the scalar field,
since the sphaleron mass is determined at tree-level from
the bosonic sector of the theory only.
If the heavy fermions do not mix with lighter fermions,
then the sphaleron process from a state with no heavy fermions
to a state with heavy fermions is prohibited strictly on the basis of
kinematics\footnote{In realistic models with fermionic mixing,
the rate of sphaleron decay will be suppressed by a
Cabbibo angle which could be quite small for heavy fermions.}.
Setting aside for the moment the obvious technical problems
accompanying a large Yukawa coupling,
this simple scenario illustrates that fermions with
large Yukawa couplings to scalars are expected to modify
the scalar background field\cite{BAG}.

One can imagine at least two ways in which the scalar
field could behave under the influence of heavy fermions.
It may be that the sphaleron is stabilized by the effects
of heavy fermions,
if its negative modes are turned positive by fermion loop effects
\footnote{A situation like this has been considered recently
in an investigation of sphaleron stability in the presence of
zero-modes\cite{VAC}.}.
It may also be that the sphaleron mass is simply boosted by
effects of heavy fermions, but remains an unstable saddlepoint.
This latter effect will be demonstrated in the following pages.

The sphaleron rate in thermal equilibrium,  $\Gamma $,
 is controlled by the
Boltzmann exponential \cite{KUZ}
\beq
\Gamma \, \propto \,
C\,\times\, \exp{\left[-\beta\, E_{sph}\right]} \,
\eeq
for temperatures $T \equiv 1/\beta $ below the mass of the sphaleron,
$E_{sph}$.
The sphaleron is a solution of the time-independent classical
equations of motion for scalar and gauge fields determined
by minimizing the energy functional.
In the one-loop approximation,
the prefactor $C$ is determined by calculating the gaussian and zero-mode
fluctuations around the sphaleron.
This factor represents the entropy associated with the sphaleron,
and it combines with the energy to give the free energy of the sphaleron.
However,
this free energy will not in general be a minimum of the
free energy functional.
The self-consistent method for determining the sphaleron rate is
to first calculate the free energy functional and then
minimize it.
This determines a new configuration, the {\it thermal sphaleron}.
The rate is then given by the equation
\beq
\Gamma \, \propto \, \exp{\left[-\beta\, F_{sph}\right]} \, .
\eeq
This rate will include all of the effects of fermionic fluctuations.

In this paper such a  self-consistent procedure is performed in a simple model,
in the limits where the fermion mass is greater than or less
than the temperature\cite{WAS,NAC}.
The fermionic determinant is found for an arbitrary
scalar background, using a WKB approximation,
so that the effects of gradients to all orders are included.
Thus, the problems involved in non-locality mentioned above
may be addressed.

In the high $T$ limit,
it is shown that the finite temperature effects serve
to make the gradient expansion well-defined, as expected.
This allows a comparison of the WKB result with the gradient expansion.
The fermion mass is effectively controlled by the temperature,
and the gradients are suppressed by inverse powers of this dimensionful
quantity.
However,
the fermions do not decouple in the high $T$ limit,
since they contribute to temperature-dependent renormalizations
of the parameters in the theory which are physically observable.
In the simple model considered here,
the only effect of the fermions in this limit is a temperature-dependent
renormalization of the scalar mass.
In the low $T$ limit,
it is shown that the heavy fermions also do not decouple, however
their effects cannot be summarized by a temperature-dependent
renormalization of the bare parameters.
Furthermore, it is shown that the gradients are still suppressed
in the regime where quantum fluctuations become important.

The rest of the paper is structured as follows.
Section 2 contains a review of the 1+1 Abelian Higgs model
which has many of the features of the electroweak theory
necessary for a study of finite temperature fermion number violation.
In Section 3,
the finite temperature effective action for the 1+1 Abelian Higgs model
is calculated to all orders in gradients, using a WKB approximation.
The details of these calculations are lef tfor the appendix.
Section 4 contains a discussion of the contribution of heavy fermions
to the free energy,
and calculations of the thermal sphaleron in high and low temperature regimes.
Finally, the last section contains a discussion of the results.

\section{The Model}

The 1+1 dimensional Abelian Higgs Model, chirally coupled
to fermions, contains several features necessary for a study
of finite temperature fermion number violation
\cite{CAR,BOC,DIN}.
The Minkowski action of the model is given by:
\beqa
\label{eq:action}
S & = &
N \int \, d^{\, 2} x \;~
- {1\over 4}\, F^2_{\mu\nu} \: + \: \vert\, D_{\mu}\Phi\,\vert^2
\: - \: \lambda\,\left(\, \vert\Phi\vert^2 \: - \: v^2/2\,\right) \\
  &   &
+ \:
\sum_a^N \: \int \, d^{\, 2} x \;
\bar{\psi}^a_L\, i D_L\,\psi^a_L \: + \: \bar{\psi}^a_R\, i D_L\,\psi^a_R
\: + \:
g\,\bar{\psi}^a_L\,\Phi\,\psi^a_R \, +\,
g^*\,\bar{\psi}^a_R\,\Phi^*\,\psi^a_L \, . \nonumber
\eeqa
The $U(1)$ gauge field $A_\mu$ is chirally coupled to $N$ flavors of
fermions through the covariant derivative,
\mbox{$D_{L,R} \equiv \gamma^{\mu} (\partial_{\mu} \pm i e A_{\mu})$ }.
The complex scalar $\Phi $ is responsible for spontaneously
breaking the $U(1)$ symmetry.
The dimensionful parameters in the theory are $\lambda$ with
dimensions of mass-squared, and $e$ and $g$ with dimensions of mass.
The scalar mass is $M_\phi = \sqrt{2\lambda}v$ and
the fermion mass is $m_f = gv$ \footnote{CP violating phases will be ignored.}.
By a rescaling of the scalar field,
it can be shown that the limit of weak scalar self-coupling corresponds
to $v >> 1$.

The work presented in the following sections is based on
a large $N$ expansion of the free energy of this model so that
the effects of a large Yukawa coupling can be considered\cite{BAG,NAC}.
An overall $N$ has been included in the bosonic part
of eq.\ref{eq:action} to make this expansion well-defined
\footnote{An action with canonically normalized kinetic terms
but rescaled parameters is obtained by rescaling the bosonic fields
by $1/\sqrt{N}$.}.
The large $N$ limit is taken with the parameters $e$, $g$,
$\lambda $ and $v$ in eq.\ref{eq:action} held fixed.
Then,
the tree-level bosonic action and the one-loop fermion determinant
are the leading-order $O(N)$ contributions to the free energy.
Bosonic loops are suppressed by powers of $1/N$ and are neglected.

The model contains a conserved gauge current,
$
J_{\mu} = \sum_a \,
\bar{\psi}^a_{L} \gamma_{\mu} \psi^a_{L} -
\bar{\psi}^a_{R} \gamma_{\mu} \psi^a_{R} ,
$
and global fermion number currents associated with each fermion flavor,
$
J^a_{\mu} =
\bar{\psi}^a_{L} \gamma_{\mu} \psi^a_{L} +
\bar{\psi}^a_{R} \gamma_{\mu} \psi^a_{R} ,
$
which obey the anomalous conservation equation
\beq
\label{eq:anomaly}
\partial^\mu J^a_\mu \: = \:
{e\over 4\pi}\,\epsilon^{\mu\nu}\, F_{\mu\nu} \hspace{5mm} \forall\: a \, .
\eeq
As in the electroweak theory,
the anomaly appears in the divergence of a vector current
due to the asymmetry in couplings between left- and right-handed
fermions to the gauge field.

The vacuum structure of this theory is non-trivial and is
described by the homotopy group, $\pi_1 (S^1)\, = \, {\bf Z}$,
as it is in the electroweak theory.
Each inequivalent vacuum is labelled by the Chern-Simons number:
\beq
\label{eq:ChernSimons}
N_{CS} \: \equiv \:
{e\over 2\pi}\,\int d\, x\: \epsilon_{0\nu}\, A^{\nu}
\eeq
which takes values in ${\bf Z}$.
Imposing periodic boundary conditions in a box of size $L$,
the classical vacuum with topological number $N_{CS}$ is given by
\beq
\label{eq:vac}
\Phi \, = \, {v\over \sqrt{2}}\: e^{i\alpha(x)} \hspace{5mm} , \hspace{5mm}
A_\mu \, = \, {1\over e}\,\partial_\mu \alpha(x),
\eeq
where $\alpha(0) - \alpha(L) = 2\pi\, N_{CS}$ and $N_{CS}\,\in\, {\bf Z}$.
Working in the temporal gauge $A_0 = 0$,
the Coulomb gauge condition $\partial_1 A_1 = 0$ fixes the remaining
time-independent gauge freedom and eq.\ref{eq:ChernSimons}
determines the vacuum gauge
field to be $\{ A_0 = 0 , A_1 = 2\,\pi\, N_{CS}/e L \}$.

Fermion number violation occurs for each fermion flavor
when the spacetime integral of eq.\ref{eq:anomaly}is non-vanishing.
This is the case for gauge fields with non-vanishing winding number,
defined as
\beq
\label{eq:winding}
\nu \: \equiv \:
{e\over 4\pi}\,\int d^{\, 2}\, x\:\epsilon_{\mu\nu}\, F^{\mu\nu} \, .
\eeq
Configurations with non-zero winding number describe transitions
between adjacent vacuua since
\beq
\label{eq:change}
N_{CS} (t = +\infty ) - N_{CS} (t = -\infty ) \: =\:
{e\over 2\pi}\,\int d\, t\, d\, x\: {\partial\over\partial t}\,
\epsilon_{0\nu}\, A^{\nu}  \: = \: \nu .
\eeq
Transitions between adjacent vacuua, whether by tunnelling or by
finite temperature fluctuations,
are necessarily accompanied by fermion number violation,
according to eq.\ref{eq:anomaly} and eq.\ref{eq:winding}.
At zero temperature,
the Abrikosov-Nielsen-Olesen vortices are Euclidean
configurations, or instantons, which describe such tunnelling events\cite{ANO}.
At finite temperature,
sphalerons, to be discussed below, are assumed to be the relevant
configurations for transitions between adjacent vacuua.

In the absence of fermions,
it is well-known that the 1+1 ABH model does not exist in the Higgs phase.
Instantons destroy the long range order and a confining
phase arises \cite{COL}.
However,
the existence of fermion zero modes in the instanton background
implies the vanishing of instanton contributions to
fermion-number-conserving Greens functions\cite{KRA}.
In particular,
the effective potential which determines the ground state of the
theory receives no contribution from instantons,
and there is no restoration of symmetry when fermions are included in
the model\cite{KRA,RU}.
Therefore, this model is expected to exhibit the symmetry-breaking
implied by the tree level potential.


The 1+1 ABH theory contains a non-contractible loop in the
configuration space of the bosonic sector of the theory
which interpolates between vacuua differing by one unit of $N_{CS}$\cite{BOC}.
The field configuration of maximum energy on this loop
is the {\it sphaleron},
which determines the height of the free energy barrier between
adjacent vacuua in the broken phase of the theory.
The gauge field on the non-contractible loop has
the vacuum form $\{ A_0 = 0 , A_1 = 2\,\pi\, N_{CS}(\tau )/e L \}$.

The scalar field on the non-contractible loop is
now simply described by a scalar field in the form
$\Phi' (x) \equiv e^{-2\pi i N_{CS}(\tau)\, x/L}\, \Phi (x)$,
which removes the dependence on the constant gauge field\cite{BOC}.
Identifying the Chern-Simons number on the loop, $N_{CS} (\tau )$,
with the loop parameter itself $\tau = \left[0, 1\right]$,
and solving the time-independent equations of motion then gives
the scalar field :
\beq
\label{eq:NCL}
Re\,\Phi'\, (x) \, =\, {\tilde{v}\over \sqrt{2}}\, \tanh
\sqrt{{\lambda\,\tilde{v}^2\over 2}}\, x \, ,
\hspace{1.0cm}
Im\,\Phi'\, (x) \, =\, {v\over \sqrt{2}}\, \cos \pi\,\tau \, ,
\eeq
where $\tilde{v}^2 \equiv v^2\,\sin^2\,\pi\tau $.
By convention,
the Imaginary part of $\Phi'$ has been chosen to be spatially constant.
So, the Real part of $\Phi'$ interacts in a ``slice'' of the
Higgs potential, which is a double-well potential.
The solution is then a familiar ``kink'' configuration of the theory
of a single scalar field interacting through a double-well potential,
with a mass controlled by the loop parameter $\tau$.

The energy functional reduces to
$E (\tau ) = {2\sqrt{2}\over 3}\,\sqrt{\lambda}\,
v^3\,\vert\,\sin^3\,\pi\,\tau\,\vert$
for configurations on this loop.
The energy reaches a maximum when $\tau = 1/2$,
and this defines the unstable sphaleron configuration.
In terms of the original variables, it takes the form:
\beq
\label{eq:sphaleron}
\Phi \, = \, {v\over\sqrt{2}}\,\tanh\left(\,\sqrt{{\lambda\, v^2 \over 2}}
\, x\,\right) \,
\exp{\left(\,i\,\pi\, x/L\,\right)} \: , \hspace{5mm}
A_1 \, = \,  {\pi\,\over e\, L} \: , \hspace{5mm}
A_0 \, = \, 0 \, .
\eeq
This is an unstable static solution to the source-free equations
of motion for the scalar field.
The $\tau$-direction of the energy functional along the loop
represents the only unstable direction in configuration space away
from the sphaleron.
The sphaleron energy,
$E_{sph} = {2\sqrt{2}\over 3}\,\sqrt{\lambda}\, v^3 =
{2\over 3}\,v^2 \, M_\phi $,
has the familiar form of a non-perturbative result since the
weak coupling regime is achieved for $v >> 1$
\footnote{Analogously, the electroweak sphaleron has energy,
$E_{sph} \propto M_w/\alpha_w$.}.
In the following section,
the contribution of fermions to the free energy of this system
will be calculated to determine thermal and quantum corrections to
this classical sphaleron configuration.
The constant gauge field in eq.\ref{eq:sphaleron}
does not contribute to the fermion determinant to be calculated,
and so the gauge field will be neglected below\cite{BOC}.
In this sense,
the approach in this paper is complimentary to \cite{BOC}
where the Yukawa coupling is neglected but bosonic fluctuations
around the sphaleron are calculated.

The sphaleron plays a key role in determining the
rate of fermion number violation in the theory.
As determined above,
the sphaleron energy is the height of the free energy barrier
between adjacent vacuua.
If the sphaleron is also assumed to be the unique saddlepoint of
{\it lowest} free energy,
then its energy controls the rate of finite temperature fluctuations
between adjacent vacuua.
As mentioned in the Introduction,
this rate is accompanied by a Boltzmann suppression factor
\beq
\Gamma\propto\exp{\left[\, -\,\beta \, E_{sph}\,\right]}
\eeq
for temperatures much lower than the barrier height
but much larger than the scalar mass,  $M_\phi << T << v^2 M_\phi $.
Similarly,
the rate of fermion number violation which accompanies vacuum transitions
is also determined by the Boltzmann exponential in this temperature
range\cite{CKN}.
The strong dependence of the rate on the energy of the sphaleron
requires an accurate determination of this quantity and provides
further motivation for the work in this paper.

\section{The Free Energy in the WKB Approximation}

The partition function for the system of
fermions in a background scalar field, $\Phi$,
can be represented as a Euclidean path integral,
\beq
\label{eq:partition}
Z \left[ \Phi \right] \, =\,
\exp\left[
-N\ln\, det\left(i\not\! \partial_E  - gv\right)
\right] \,
\int iD \psi D \psi^{\dagger} \,
\exp\left[
-\int_0^\beta \, dt\,\int d^3 x \:
{\cal L}_E + \delta {\cal L}
\right] \, .
\eeq
The fermion fields $\psi$ and $\psi^\dagger $ are
required to be anti-periodic with period $\beta\equiv 1/T$.
$ {\cal L}_E$ is the Wick rotation
of the  Lagrangian given in eq.\ref{eq:action},
and
$\delta {\cal L }$ is the counterterm Lagrangian.
The determinant factor is the finite temperature determinant in the constant
background which serves to normalize the free energy.
The counterterm Lagrangian $\delta {\cal L}$ is given by
\beq
\label{eq:counter} \delta {\cal L}=AN( \vert\Phi\vert^2 - v^2/2 ),
\eeq
and is sufficient to renormalize both the one point and the two point
function.
$A$ is chosen so that the tadpole graph  vanishes when $\Phi=v$.
This prescription gives
\beq
\label{eq:counterterm}
A \, = \, -{g^2 \over 2\pi} \, \int^\infty_0
{dp\over \sqrt{p^2 + g^2 v^2}} \, .
\eeq
Using this value for $A$,
the physical mass will be
\beq
\label{eq:mphys}
m^2_{phys}=N\, (-2\lambda v^2-g^2/\pi).
\eeq
This
counterterm is sufficient to renormalize the theory
at finite temperatures.

When the gaussian integral over the fermionic fields is performed,
the partition function is expressed as the determinant of the operator
\beq
\label{eq:lnZ}
\ln Z \, = \, {N\over{4}} \, \sum_\epsilon \sum_n \,
\ln \left( \omega_n^2 + \epsilon^2 \right) \, ,
\eeq
where $\omega_n$ are the Matsubara frequencies given by
$\omega_n=2\pi (2n+1)/\beta $ and $\epsilon$ are the
eigenvalues
of the spatial operator $O$ given by
\beq
\label{eq:operator}
O \, = \,
\gamma_0^E \, \left[ \,
i\not\!{\partial} \, - \,
\lambda \phi P_L \, - \, \lambda^* \phi^* P_R
\,\right] \, .
\eeq
$P_L$ and $P_R$ are the projection operators on the left and right moving
states, respectively.

The free energy of this system,
defined by the usual thermodynamic relation,
\mbox{$F \, = \, -{1\over\beta}\,\ln Z$,}
is an effective action for the scalar field, \mbox{$\phi$.}
The remainder of the paper will focus on a calculation of this
quantity using the WKB approximation.

Performing the sum over the Matsubara frequencies in eq.\ref{eq:lnZ}
leaves \footnote{For notational convenience we don not include here the
ideal gas contribution}
\beq
\label{eq:det}
F \, = \, - {N\over 2}\, \sum_\sigma\, \sum_\epsilon \,
\left[\, f (\epsilon ) - f (\epsilon_0 ) \,\right] \, .
\eeq
where $f (\epsilon )$ is the contribution to the free energy
from states with energy $\epsilon$
\beq
\label{eq:fe}
f (\epsilon ) \, \equiv \, \epsilon \, +\,
{2 \over \beta} \, \ln \left( 1 + e^{-\beta \epsilon} \right) \, .
\eeq
The sum over $\sigma$ corresponds to the sum over two helicity states,
and $\epsilon_0$ are the eigenvalues of the operator $O$ with
$\phi$ taking on the constant value $v$.
Eq.\ref{eq:det} reduces to the shift in the Dirac sea in the
zero temperature limit\cite{NAC}.

The calculation of the free energy has been recast as a
calculation of the spectrum of the operator $O$.
In terms of the up (+) and down (-) components of the
wave function,
the eigenvalue problem may be expressed as a Schr\"{o}dinger equation,
\beq \label{eq:scrd}
{d^2\psi_{\pm}\over{dx^2}}
\, - \,
\left( \,
R^2 \, + \, I^2 \, \mp \, {dR \over dx} \, + \,
{dI \over dx} \, \sigma_1 \,
\,\right)\, \psi_{\pm}
\, = \, - E_{\pm}^2 \, \psi_{\pm} \, .
\eeq
A convenient basis for the two dimensional Dirac matrices has been
chosen in which \mbox{$\gamma_0=\sigma_1 $ ,} \mbox{$\gamma_1=i\sigma_3 $ ,}
and \mbox{$\gamma_5=\sigma_2$.}
The scalar field has been expressed here in terms of
\mbox{$R \equiv Re(g\,\Phi)$}
and \mbox{$I \equiv Im(g\,\Phi)$ }
\footnote{We will assume that there are no CP violating phases.}.

The imaginary part of the scalar field will be assumed to play
no role in the energetics of the sphaleron.
This is clearly seen in the temporal gauge
where the imaginary piece is utilized only to enforce
the boundary conditions at $x = \pm \infty $.
The gradient energy of the phase in eq.\ref{eq:sphaleron} will
vanish in the infinite volume limit.
Thus,
the eigenvalue equations eq.\ref{eq:scrd} can be decoupled in the
temporal gauge for the general sphaleron-like background.

It is convenient to follow the notation used in ref.\cite{NAC}
and introduce the parameters
\beq
\label{eq:defs}
x_{cl} \, =\, \sqrt{{2\over \lambda v^2}} \hspace{0.2cm} , \hspace{0.2cm}
y \, = \, {g\over\sqrt{\lambda}} \hspace{0.2cm} , \hspace{0.2cm}
z \, = \, x/x_{cl} \, .
\eeq
The Compton wavelength of the scalar, $x_{cl}$, and the
inverse temperature, $\beta$, are now the only dimensionful parameters.
The size of the classical sphaleron is roughly of order $x_{cl}$.
The parameter, $y$, is the ratio of the scalar and fermion Compton
wavelengths.
Now the eigenvalue equation may be written in the simple form:
\beq
\label{eq:schrodinger}
\left[
{d^2\over{dz^2}}-y^2V_{\pm}(z)-y^2+x_{cl}^2\epsilon^2_{\pm}
\right]
\,\psi_{\pm} \, = \, 0 \, ,
\eeq
with a ``potential'' given by
\beq
V_{\pm}(z) \, = \,
\phi^2/v^2-1 \, \mp \, {1\over{yv}} \,
\left( {d \phi \over d z} \right) \, .
\eeq
in terms of a canonically-normalized Real scalar field,
$\phi\equiv \sqrt{2}\, Re(\Phi ) $.

The eigenvalue problem will now be solved using a WKB approximation
\cite{WAS,NAC}.
This approximation is expected to be valid when the scalar field
varies on scales larger than the fermion Compton wavelength,
or roughly for large $y$.
Since the goal is to study the effects of fermions on the sphaleron,
only background scalar fields with the same topology as the sphaleron
will be considered.
These consist of fields which interpolate between two minima of
the potential as $x$ varies from $-\infty$ to $+\infty$,
as in eq.\ref{eq:sphaleron}.
For these fields,
the potential $V$ will be parity even, and
the eigenstates will be classified by their parity quantum number.
Furthermore,
these eigenstates naturally fall into two classes:
continuum states and discrete (bound) states.

\subsection{The Continuum States}

The asymptotic form of the WKB wavefunctions will be given by
plane waves
\beq
\label{eq:evenwfns}
\psi_\sigma^{even}(k,z) \: \rightarrow \:
\cos\left( kz\pm {1 \over 2} \delta_\sigma^{even} \right) \; ,
\hspace{0.3cm}
\psi_\sigma^{odd}(k,z) \: \rightarrow \:
\sin\left( kz\pm {1 \over 2} \delta_\sigma^{odd} \right)
\; , \hspace{0.3cm}
z \rightarrow\infty \, .
\eeq
The momenta here are given by the free dispersion relation:
\beq
\label{eq:momentum}
k \, = \,\sqrt {x_{cl}^2 \epsilon^2 -y^2} \, .
\eeq

Periodic boundary conditions in a finite box of spatial size $L$
require that the WKB phase shifts $\delta_{\sigma}$ be related
to the momenta by: $k_{\sigma n}^0L + \delta_{\sigma}(k) = 2\pi n$.
As the size of the box is eventually to be taken infinite,
eq.\ref{eq:det} can be expanded in powers of $\delta/L$,
leaving
\beq
\label{eq:taylor}
F \, = \, - {N\over 2}\, \sum_\sigma\, \sum_k \,
\left[\,
-\left(\, {\partial f \over \partial \epsilon } \,
{\partial \epsilon \over \partial k } \,\rule[-3mm]{0.2mm}{9mm}_{\, k_0}
\,\right) \,
\left(\, {\delta_\sigma\over L} \,\right)
\, + \, O\left(\, {\delta_\sigma^2\over L^2} \,\right)
\,\right] \, ,
\eeq
where the phase shifts for even and odd parity wavefunctions
have been combined,
\beq
\delta_{\sigma} \, \equiv \,
{1\over2} \, \left(\,\delta^{even}_\sigma \, + \,
\delta^{odd}_\sigma \,\right) \, .
\eeq

The phase shifts contain all of the information about the
potential and are given by the WKB approximation.
\beq
\label{eq:phase}
\delta_\sigma ( k ) \, = \, \int^\infty_{- \infty}dz \,
\left[\, k_\sigma(z) - k \,\right] \hspace{0.4cm} , \hspace{0.4cm}
k_{\sigma}(z) = \sqrt{x_{cl}^2\epsilon^2 - y^2 - y^2V_{\sigma}(z)}.
\eeq

In the limit of large $L$,
the sum in eq.\ref{eq:taylor} becomes an integral
and the contribution of continuum states to the free energy is:
\beq
\label{eq:master}
F_{cont} \, = \, {N\over 2\pi x_{cl}} \,
\sum_\sigma\, \int^\infty_0 dk \,
{k \over \sqrt{k^2+y^2} } \,
\left[\, 1 \, - \, {2 \over 1 + e^{\beta\epsilon (k)}} \,\right] \,
\delta_\sigma (k) \, ,
\eeq
with $\epsilon$ given in terms of $k$ by the free particle
dispersion relation, eq.\ref{eq:momentum}

The first term in the square brackets in eq.\ref{eq:master}
is the zero temperature contribution.
It is UV divergent as it stands and must be regulated,
after which the momentum integral can be carried out explicitly.
The counterterm discussed previously, eq.\ref{eq:counterterm},
renormalizes this term and leaves\cite{NAC}
\beqa
\label {eq:zerotcont}
F^0_{cont} & = &
 -\, {y^2 N \over 4\pi x_{cl}} \,\sum_{\sigma} \,
\int^\infty_{-\infty} dz \:
\left[ \:
{1\over 2} \, \left(1- {\phi^2 \over v^2 }\right) \: - \:
\sqrt{-V_\sigma}  \right. \hspace{2cm}  \\
&   & \hspace{6cm} \left.
+ \: \left( \, 1 + V_\sigma \, \right) \,
\ln \left(\, 1 + \sqrt{ - V_\sigma } \,\right)
\:\right] \, . \nonumber
\eeqa

The second term in the square brackets in eq.\ref{eq:master}
is the finite temperature contribution and may be written
in terms of the ratio of temperature to fermion mass,
$a \equiv \beta y/x_{cl} $.
\beq
\label{eq:cont}
F (a^2)_{cont} = - \, {N x_{cl} \over \pi \beta^2} \,
\sum_\sigma \int^\infty_{- \infty} dz \,
\int^{\infty}_0 \, {dx \, x \over \sqrt{ x^2 + a^2 } } \,
{1 \over 1 + e^{ \sqrt{x^2 + a^2} } } \,
\left\{\, \sqrt{ x^2 - a^2 V(z) } - \sqrt {x^2} \,\right\}
\eeq
This integral is UV finite,
but cannot be evaluated analytically for arbitrary values
of the temperature-to-mass ratio, $a$.
However, its limiting forms can be.
The finite temperature contributions will be evaluated
in the limit of large and small $a^2$ below.
But first,
the contribution of discrete states to the free energy
must be included.

\subsection{The Discrete States}

The spectrum of discrete states is determined
in the WKB approximation by the Bohr-Sommerfeld condition.
The Schr\"{o}dinger equation, eq.\ref{eq:scrd},
will have a discrete bound state whenever
\beq
w_\sigma (\epsilon ) \, \equiv \,
{1\over{\pi}} \int^{\infty}_{- \infty} dz \:
k_{\sigma}(z) \,
\Theta \left( k^2_\sigma(z) \right) \, .
\eeq
equals half of an odd integer.
The total number of discrete states is then given by the
integer closest to $w_\sigma (y/x_{cl})$.
Therefore,
the contribution to the free energy, eq.\ref{eq:det},
from the discrete states will be given by\cite{NAC}
\beq
\label{eq:discstart}
F_{disc} \: = \:  - {N\over 2} \, \sum_\sigma \,
\sum_{w_\sigma > 0}^{w_\sigma (y/x_{cl})} \:
f \left( \epsilon\right) \, - \,  f \left( y/x_{cl}\right) \, .
\eeq

This sum may always be expressed in terms of an integral
\beq
\label{eq:intapprox}
F_{disc} = - {N\over{2}} \,
\sum_\sigma \,
\int^{w(y/x_{cl})}_0 dw \:
\left[\,
f \left( \epsilon\right) \, - \,  f \left( y/x_{cl}\right)
\,\right]
\: + \: Rem.
\eeq
with a remainder given by the Euler-MacLaurin formula.
The remainder represents the error in the free energy from
an integral approximation.
As the ``potential'' becomes deeper,
the number of bounds states $w (y/x_{cl}) $ increases,
and the integral approximation becomes better.
The depth of the potential is controlled by the dimensionless
parameter $y$, the ratio of the fermion and scalar Compton
wavelengths.
Thus,
the integral approximation is obtained in the WKB regime,
where the scalar field is slowly varying on the scales of the
fermion Compton wavelength.
This approximation will be discussed further below.

Finally,
eq.\ref{eq:intapprox} may be cast in the same form as
the contribution from the continuum states, eq.\ref{eq:cont},
by changing variables from $w$ to $\epsilon$.
\beq
\label{eq:intapproxchanged}
F_{disc} \: = \: {N\over 2\pi } \,
\sum_\sigma \,
\int_{-\infty}^{\infty} dz \,
\int^{y/x_{cl}}_{y\sqrt{1 + V_\sigma}/x_{cl}} d\epsilon \,
\left[\,
1 - {2 \over 1 + e^{\beta\epsilon}}
\,\right] \,
\sqrt{x_{cl}^2\epsilon^2 - y^2(1 + V_\sigma)}
\eeq

The first term in the square brackets is the zero temperature
contribution.
The integral over $\epsilon$ can be done explicitly to give:
\beq
\label {eq:zerotdisc}
F^0_{disc} \: = \: -\, {y^2 N \over 4\pi x_{cl}} \,
\sum_{\sigma} \, \int^\infty_{- \infty} dz \:
\left[ \,
\sqrt{-V_\sigma} \: + \:
\left( \, 1 + V_\sigma \, \right) \,
\ln \left(\,
{ \sqrt{1 + V_\sigma} \over {1 + \sqrt{- V_\sigma}} }
\,\right)
\,\right] \, .
\eeq

The second term in the square brackets is the finite temperature
contribution and may be written in terms $a$, defined previously.
Changing variables to $u \equiv \beta\epsilon$ gives:
\beq
\label{eq:disc}
F(a^2)_{disc} \: = \:
{N\over \pi } \, \sum_\sigma \,
\int_{-\infty}^{\infty} dz \,
\int^a_{a\sqrt{1 + V_\sigma}} \, {du \over {1 + e^u}} \,
\sqrt{u^2 - a^2(1 + V_\sigma)}
\eeq
Finally,
the sum of the two zero temperature pieces,
eq.\ref{eq:zerotcont} and eq.\ref{eq:zerotdisc},
and the two finite temperature pieces, eq.\ref{eq:cont}
and eq.\ref{eq:disc}, yields the total free energy, eq.\ref{eq:det}.

\subsection{The High Temperature Expansion}

The high temperature limit will provided a check on the
previous expressions.
One expects the leading finite temperature effects
to be summarized in a temperature-dependent mass term
for the scalar field.
The high temperature expansion is defined as $a^2 << 1$.
The free energy is not an analytic function of the temperature
in this limit,
and special care must be taken to obtain the expansion for
small $a$ (see Appendix).
Including only the first non-trivial dependence on $a^2$,
eq.\ref{eq:cont} is:
\beqa
\label{eq:Fintapprox}
F (a^2)_{cont}  & = &
{y^2N\over{2\pi x_{cl}}} \,
\sum_{\sigma} \,
\int^{\infty}_{-\infty} dz \,
\left[ \, \rule{0cm}{0.7cm}
{1 \over 2} \sqrt{ - V_{\sigma}(z) } \, -  \,
{1 \over 4} \left(\, 1 + V_{\sigma} (z)  \,\right) \,
\left( \,
2\,\gamma \, - \, \ln \left(a^2/\pi^2 \right)
\, \right)
\right.                                       \nonumber \\
&   &
\left. \hspace{3.5cm}
- {1 \over 2 } \,
\left( \, 1 + V_{\sigma} (z) \,\right)
\ln \left( \, 1 + \sqrt{- V_{\sigma} (z) } \, \right)
\, \right]  \, + \, O(a^2)
\eeqa
When this is combined with the zero temperature contribution,
eq.\ref{eq:zerotcont},
the square root and logarithmic terms cancel precisely.
Then,
the contribution of the continuom states in the high temperature limit is:
\beq
\label{eq:contht}
F (a^2)_{cont} \: = \: {y^2N \over{4\pi x_{cl}}}
\sum_\sigma\,
\int^\infty_{-\infty} dz \,
\left[ \,
{1 \over 2} \left(\, 1 - \phi^2/v^2 \,\right)
\, - \,
{1 \over 2}\, V_\sigma (z) \,
\left(\, 2 \,\gamma - 1 + \ln \left({a^2\over\pi^2}\right) \,\right)
\, \right] \, .
\eeq
The field gradients contained in $V$ will
cancel in the sum over $\sigma $, leaving only  a
temperature-dependent mass term for the scalar field.
Thus,
it is clear that the finite temperature effects have made
the theory local, in the sense that there is a well-defined gradient expansion.

The contribution of discrete states follows from
eq.\ref{eq:intapproxchanged}.
It is not hard to demonstrate that the Euler-MacLaurin
remainder is suppressed by at least $O(a)$.
Furthermore,
since the integral in eq.\ref{eq:intapproxchanged} is cut off at the scale
$y/x_{cl}$,
the Boltzmann exponential may be expanded in the high temperature
approximation,
and the zero temperature contribution cancels out, leaving
\beq
\label{eq:dsicfree2}
F_{disc} \: = \:  \, O(a) .
\eeq
As expected, low-lying bound states do not contribute to the
free energy for high temperatures.

Combining eq.\ref{eq:contht} and eq.\ref{eq:dsicfree2}
the total free energy in the high temperature approximation
is given by
\beq
F \, = \,
{y^2 N \over 4\pi x_{cl}} \,
\left(\, 2\gamma + \ln \left({a^2\over\pi^2}\right) \,\right)
\int^\infty_{-\infty} dz \,
\left(\, 1 - \phi^2/v^2 \,\right) \, .
\eeq
We may now calculate the critical temperature in the high temperature
approximation by combining this result with the tree-level scalar
action from eq.\ref{eq:action}.
We will define critical temperature as the temperature for which
the curvature at the origin vanishes.
This is not necessarily the
temperature at which the phase transition occurs.
If the transition is first order,
then bubble nucleation will induce the phase transition at
a temperature above the critical temperature as defined here.
Therefore,
the critical temperature may {\it overestimate}
the temperature at which the transition occurs,
but this will be sufficient for this paper.
Requiring ${\partial \over \partial \phi^2}\, F_{total} = 0 $
at the origin gives
\beq
\label{eq:tc}
T_c \: = \:  \pi\, e^{-\gamma/4} \, {y\over x_{cl}} \,
\exp{\left[{2\pi v^2 \over y^2}\right]}
\, .
\eeq
Interestingly enough,
there is a range of temperature available in which
a high temperature expansion, $T >> y/x_{cl}$, is consistent
with a broken phase, $T < T_c$, and the sphaleron effects are of
interest.
So,
for at least some range of heavy fermion mass,
the leading temperature-dependent effect on the sphaleron
is summarized by introducing a temperature-dependent mass for
the scalar field,
which in turn implies a temperature-dependent sphaleron mass.
Thus,
the sphaleron mass decreases with the scalar mass as the critical
temperature is approached from below. However, notice that
in the region of validity of the high $T$ expansion
($v>>y$), it is not expected that
fermions will have a large effect on the sphaleron mass.
Contrary to the hihg $T$ case, in the low $T$ limit the
broken phase can be maintained for a wider range of
heavy fermion masses. Therefore, we would expect that in this limit
fermions will play a more important
 role in determining the sphaleron free energy.

\subsection{The Low Temperature Expansion}

The low temperature expansion is defined as $a^2 >> 1$.
It can be developed from eq.\ref{eq:cont} by changing variables from
$ x $ to $u \equiv \beta\epsilon = \sqrt{x^2 + a^2}$.
The finite temperature contribution to the free energy from
the continuous states is then:
\vspace{2mm}
\beq
\label{eq:lt}
F (a^2)_{cont} \: = \:
- \, {N x_{cl} \over \pi\beta^2} \,\sum_{\sigma} \,
\int_{-\infty}^{\infty} dz \,
\int_{a}^{\infty}\, { du \over {1 + e^u} } \,
\left\{\, \sqrt{u^2 - a^2 (1 + V_{\sigma})}
       \, - \, \sqrt{u^2 - a^2 } \,\right\} \, .
\eeq
\vspace{2mm}
In the limit that the temperature is much less than
the fermion mass,
the Boltzmann distribution is dominated by low energy states
and the exponential in the integrand dominates.
\vspace{2mm}
\beq
\label{eq:expandlt}
F (a^2)_{cont} \: = \:
-\, {N x_{cl} \over \pi\beta^2} \,\sum_{\sigma}
\int_{-\infty}^{\infty} dz \,
\int_{a}^{\infty}\, du \, e^{-u} \,
\left\{\, \sqrt{u^2 - a^2 (1 + V_\sigma)}
       \, - \, \sqrt{u^2 - a^2 } \,\right\} \, .
\eeq
\vspace{2mm}

The second term in the curly brackets is independent of $V$
and is simply expressed in terms of the MacDonald function, $K_1(a)$.
\vspace{2mm}
\beq
\label{eq:integrallt}
\int_{a}^{\infty}\, du \, e^{-u}\,\sqrt{u^2 - a^2} \, = \,
a \, K_1 ( a ) \, = \,
\sqrt{ {a\pi\over 2} }\, \left(\, 1 \, + \, O(1/a) \,\right) \, e^{-a} \,
, \hspace{5mm} a>>1 \, .
\eeq
\vspace{2mm}
The first term in the curly brackets cannot be evaluated in
terms of simple functions.
However,
an integration by parts gives the leading temperature dependence
immediately.
\vspace{2mm}
\begin{eqnarray}
\label{eq:nonvanlt}
\int_{a}^{\infty}\, du \, e^{-u}\,\sqrt{u^2 - a^2(1 + V)}
& = &
a\sqrt{-V}\, e^{-a} \, + \,
\int_{a}^{\infty}\, d\, u \, e^{-u}\,
{u \over \sqrt{u^2 - a^2(1 + V)}}  \nonumber \\
& = &
a\sqrt{-V}\left(\, 1 + O(1/a) \,\right) \, e^{-a}  \, , \hspace{5mm} a>>1 \, .
\end{eqnarray}
\vspace{2mm}
So, all contributions of continuum states to the free energy
are exponentially suppressed, $O(e^{-a})$, when the fermion mass
greatly exceeds the temperature.

The discrete states will give the dominant contribution to the
free energy in the low temperature regime.
This contribution is obtained from
from eq.\ref{eq:zerotdisc} and eq.\ref{eq:disc}.
The finite temperature part is:
\vspace{2mm}
\beq
\label{eq:disclt}
F(a^2)_{disc} \: = \:
- 2 \, {N x_{cl} \over \beta\pi} \,
\sum_{\sigma} \, \int_{-\infty}^{\infty} dz \,
\int_{a\sqrt{1 + V}}^a {du \over 1 + e^u } \,
\sqrt{u^2 - a^2 (1 + V_\sigma )} \, .
\eeq
\vspace{2mm}
For some point in the $z$-integration,
the lower bound on the $u$-integral vanishes.
Therefore,
a simple expansion of the exponential as in eq.\ref{eq:expandlt}
will not be valid at this point.
This is precisely the case where the background scalar
goes through a zero.
However,
the large factor $a >> 1$ allows $-V$ to be quite
close to unity and still allow the expansion.
One requires $a(1 + V) >> 1$ which is $-V << 1 - 1/a$.
If $-V$ is not within $1/a << 1$ of unity,
then the low temperature expansion of the exponential is valid.

If $-V$ goes to unity at some point $z_0$,
the $z$-integral can be partitioned into three regions:
\mbox{$z < z_0 - b $,} \mbox{$z_0 - b < z < z_0 + b$,}
and \mbox{$z > z_0 + b$.}
If $b$ is greater than $O(1/a)$,
then the contribution of the first and third regions is
exponentially damped since $-V$ is not near unity.
Then, the finite-T contribution to the free energy in the low temperature
regime is approximately:
\vspace{2mm}
\beq
\label{eq:FT}
F(a^2)_{disc} \: = \:
- \, 2 \, {Nx_{cl}\over \beta^2\pi} \, \sum_{\sigma} \,
\int_{z_0 - b}^{z_0 + b} dz \,
\int_{a\sqrt{1 + V}}^a \, {du \over 1 + e^u} \,
\sqrt{u^2 - a^2(1 + V_\sigma)} \, + \, O\left( e^{-a} \right) \, .
\eeq
\vspace{2mm}
The total free energy is obtained by adding the $T$-independent contributions,
eq.
   \ref{eq:zerotcont} and eq.\ref{eq:zerotdisc}.
The remaining integral contains a region of size $2b$
in which $-V$ approaches unity.
A numerical computation of the free energy in this regime will be
discussed in the next section.
The qualitative behavior of eq.\ref{eq:FT} is however clear:
the finite-T contribution to the free energy decreases as
the fermion mass increases in the low temperature regime.

\section{The Thermal Sphaleron}

As discussed in the previous sections,
the thermal sphaleron is determined by minimizing the free energy
while imposing the appropriate boundary conditions.
In this section,
this calculation will be performed in the high and low $T$ limits.

In the high $T$ regime,
the boundary conditions are temperature dependent
due to the temperature dependence of the scalar expectation value.
The scalar part of the free energy can be expressed in the form
\beq
\label{eq:newfree}
F \, = \,  {N\over x_{cl}} \,
\int^{\infty}_{-\infty}dz \,
\left[ \,
{1\over 2} \left( {\partial\phi\over \partial z} \right)^2
\: + \:
{1\over 2v^2} \,
\left(\,
\phi^2 - v^2\left( T \right)
\,\right)^2
\,\right]
\eeq
with a temperature-dependent $vev$ defined as
\beq
\label{eq:Tvev}
v^2\left( T \right) \: = \: v^2 \, - \, {y^2\over 4\pi}
\left(\,
\gamma \, + \, \ln{{a\over \pi}}
\,\right)  \, .
\eeq

Fermionic fluctuations are not expected to effect
the form of the classical sphaleron which is given by
\footnote{This was demonstrated in the case of the soliton
at zero temperature in ref.~\cite{NAC}}
\beq
\label{eq:kink}
\phi(x) \: = \: v\,\tanh\left( {x\over x_{cl}} \right) \, .
\eeq
So, only the effect on the size of the sphaleron relative to
the size of the classical sphaleron $x_{cl}$ will be considered here.
The phase in eq.\ref{eq:sphaleron} may be ignored for this purpose.
It is easily seen from eq.\ref{eq:newfree}
that the thermal sphaleron, in the high $T$ case,
is given by a simple modification of the classical solution.
One replaces the $vev$ $v$ in eq.\ref{eq:kink} by $v(T)$ determined
by eq.\ref{eq:Tvev},
including its appearance in the scalar Compton wavelength $x_{cl}$.
The free energy of the ans\"{a}tz in the high $T$ approximation is then
given by
\beq
\label{eq:tsfree}
F_{sph} \: = \: {4N\over 3 }  \, {v^2(T)\over x(T)} \: .
\eeq
where a temperature-dependent scalar Compton wavelength has been
defined for convenience
\beq
\label{eq:tmass}
x(T) \: = \: \sqrt{{2\over \lambda v^2(T)}} \: .
\eeq
This is just the classical sphaleron energy with the parameters appropriately
changed to include the finite temperature effects.
The free energy of the sphaleron is an increasing function of
the fermion mass.
Furthermore,
the free energy increases only logarithmically with the temperature
in two dimensions.
As expected,
when $v>>y$,
the thermal sphaleron reduces to the classical sphaleron because in
this regime the classical contribution from the scalar
sector overwhelms the fermionic fluctuations.
Unfortunately,
as was shown in the previous section,
the high $T$ approximation will break down in the
regime where quantum fluctuations become important (i.e. $y\gsim v$).

It is interesting at this point to compare the free energy of the
thermal sphaleron
to the free energy of the classical sphaleron
including the one-loop fermion contribution, which is given by
\beq
\label{eq:clsphE}
F_{cl} \: = \:
{4 N v^2\over 3x_{cl}} \,
\left[ \,
1  + \:
{3y^2 \over 4\pi v^2}\left( \,
\gamma \, + \, \ln{{a\over\pi}}
\,\right)
\,\right]\, .
\eeq
This free energy is indeed slightly higher than the thermal sphaleron
free energy, eq.\ref{eq:tsfree}, as expected.
Of course,
this comparison should only be made at temperatures low enough that
the temperature-dependent vev of the scalar field  is still near $v$.

In general ,
it {\it may} or {\it may not} be simple to solve for the thermal
sphaleron in the high $T$ limit.
This depends upon whether or not the corrections to the gradient
terms are suppressed by finite temperature effects.
In the Standard Model at one loop order,
the gradient corrections are suppressed because there is no
renormalization of the kinetic term \cite{LIN}.
This is also the case in the 1+1 Abelian Higgs model.
However,
in this model,
the suppression persists to {\it all} orders in the loop expansion
because only the mass and the vacuum energy require renormalization in
two dimensions.
The suppression of the corrections to the gradient terms allows one
to determine the free energy functional by calculating the fermionic
fluctuations in a constant background.
The effect of the fermions is to give the couplings in the
scalar sector temperature dependence.
The thermal sphaleron is given by the classical sphaleron with
its parameters appropriately replaced with temperature-dependent
parameters.
This is the standard procedure when working in the high $T$ limit.\cite{DOL}

In the low $T$ limit,
one is no longer bound to the regime where $v >> y$ and the effect of
fermion fluctuations may become more important.
In this regime,
it is not possible to get a closed form expression for the size
of the sphaleron.
The thermal sphaleron will be determined by using the following
ans\"{a}tz for its shape
\beq
\label{eq:ansatz}
\phi_{z_0}(x) \: = \: v\,\tanh ({z\over z_0}) \, ,\hspace{1.0cm}
z \, = \, {x\over x_{cl}},
\eeq
and minimizing the free energy with respect to the parameter, $z_0$.
The use of this ans\"{a}tz assumes that the temperature
is small enough that the scalar expectation value is still near $v$.
The free energy of this ans\"{a}tz in the low $T$ limit is given by
\beqa
\label{eq:lowtfree}
F & = &  {y^2 N \over x_{cl}}\left[\,
{2v^2\over 3y^2}\left(z_0 \, +\,  {1\over z_0}\right)
\: + \: z_0 \left( {3\over 2\pi } \, - \, {\pi\over 8}\right)
\right. \nonumber \\
&   &  \hspace{2.0cm}
+ \, \left.
{2\over{\pi a^2}} \int^\infty_{-\infty}dz
\int_{a \mid  \tanh(z/z_0) \mid\!}^a \,
{du \over e^u + 1} \,
\sqrt{u^2 - a^2 \tanh^2(z/z_0)}
\,\right] \, .
\eeqa
The first term in this equation is the classical contribution,
while the second term is the zero temperature contribution from
quantum corrections.
In the $T=0$ limit,
the size of the sphaleron which minimizes eq.\ref{eq:lowtfree} is \cite{NAC}
\beq
\label{eq:notsize}
z_0 \: = \:
\left[ \,
{3 y^2 \over 2 v^2}
\left( \, {3\over 2\pi } \, - \, {\pi\over 8} \, + \,
{2 v^2 \over 3 y^2 }\, \right)
\, \right]^{-1/2}.
\eeq
Note that at zero T ,for the choice $v/y=.1$, the sphaleron size changes by a
factor
of $1/4$ relative to its classical value.
Thus the effect of {\it quantum} fluctuations alone
will tend to increase the sphaleron mass as the fermion is made heavier.
Furthermore,
the temperature-dependent effects induced by
the fermions are expected to become less important as the fermion mass
increases, because thermal fluctuations will be suppressed.

As in the high $T$ case, we expect
the classical contribution to dominate the contribution
from fermionic fluctuations in the limit that $v>>y$ and the size of
the sphaleron will reduce to the classical value $x_{cl}$, or $z_0 = 1$.
However,
allowing $y$ to become large while holding the temperature fixed,
fermionic fluctuations become increasingly important,
and there are non-negligible temperature-dependent corrections.
The free energy was calculated numerically for several values of $z_0$ and $a$.
The thermal sphaleron size was determined by minimizing
the free energy with respect to $z_0$, for a given value of $a$.
Figure 1 shows the thermal sphaleron size in units of the $T=0$ sphaleron
mass, as a function of the parameter $a$ for $v/y=0.1$.
This figure shows that in addition to the quantum fluctuations  $thermal$
fluctuations are non-negligible (yet smaller than the quantum effects)
as the temperature is increased.
Thus,
fermionic fluctuations will tend to increase the free energy of
the sphaleron relative to its classical value.
Figure 2 shows the free energy of
the thermal sphaleron in units of the classical sphaleron energy
for several values of
the parameter $a$, with the choice $v/y=0.1$.
Figure 2 indicates that quantum corrections may change the
sphaleron free energy by a factor of 5.
Furthermore, within the validity of the low $T$ expansion,
the temperature-dependent effects are at the level of 20 percent.

\section{Discussion}

The $1+1$ Abelian Higgs model has illustrated the effects of
heavy fermions on a non-perturbative scalar background.
In both the low and high temperature regimes,
a {\it self-consistent} calculation of the free energy of the
sphaleron has shown that heavy fermions will increase the mass
of the thermal sphaleron.
It has been shown that the thermal sphaleron has a smaller free energy
than the sum of the classical sphaleron energy and its fermionic corrections.
That is to say, the classical sphaleron is not a true saddlepoint
of the free energy functional.

Beyond the implications this result might have for phenomenology,
the results obtained here also address the interesting
theoretical questions mentioned in the Introduction.
In the low temperature regime,
the free energy was found to be a non-local functional of
the scalar field.
The free energy is correctly expressed in this regime only by
the summation of gradients to all orders,
a summation provided by the WKB approximation.
However,
the net effect of the fluctuation-induced gradients is suppressed
exactly when one would expect quantum fluctuations become important
($y\gsim v$).
An heuristic explanation is that the sphaleron size acts as a cut-off
for infra-red divergences.
Since the sphaleron gets
smaller
when the
fermionic mass
($y$)  increases,
the infrared cut-off increases, and fluctuation-induced gradients will
be further suppressed.
Of course, as the size of the sphaleron decreases,
the kinetic term in the classical Lagrangian
will become more important.
Though the effects of gradients are small,
thermal effects can still change the sphaleron free energy substantially.
This is a consequence of the occupation of a large number of low-lying
fermionic states which become available when the Yukawa coupling is increased.
The $T=0$ quantum effects can also change the sphaleron free energy
substantially as a consequence of the standard perturbative non-decoupling
when the fermion mass is taken to infinity while holding the scalar vacuum
expectation value constant.

In the high temperature regime on the other hand,
it has been shown here that the gradient expansion is
well-behaved.
So, the net effect of thermal fluctuations at high $T$ effects is
the localization of the theory.
As expected,
the fermions do not decouple in this regime,
even though their masses are determined by the dimensionful parameter, $T$,
since they contribute to temperature-dependent renormalizations of the
parameters in the theory which are physically observable.

Thus,
one may conclude that, in the simple model studied in this paper,
the fermions do not decouple in the {\it perturbative} sense
and can enhance the sphaleron mass.
Since the gradients are suppressed,
one may also conclude that the effects of the {\it non-perturbative}
variation of the field discussed in the Introduction are small.
One would expect similiar behaviour also in four dimensions
for the physical reasons discussed above.
The fact that there is only {\it perturbative} non-decoupling also
encourages one to expect that non-perturbative methods may not be
necessary in order to calculate the effects of heavy fermions on
non-perturbative scalar backgrounds in four dimensions.

\section{Appendix}

The integral, eq.\ref{eq:cont}, is not an analytic function
of $a $ in a neighborhood of $a = 0$.
So, a naive Taylor expansion in the region, $a \simeq 0$,
is bound to fail.
This is a familiar problem which occurs
in the calculation of the free energy of a gas of free particles.
The work in this Appendix will extract the leading non-analytic
behavior near $a \simeq 0$.

Consider the integral:
\beq
\label{eq:I}
I\, (\, a^2\, ;\, V\, ) \: =\: \int_0^{\infty}\, d\, x^2\,
\sqrt{ {x^2 - a^2 V\over x^2 + a^2} }
\:
{1\over e^{\sqrt{x^2 + a^2}} + 1}
\eeq
where $x\equiv k\beta/x_{cl}$ and
$a^2 \equiv \beta^2 m^2 = \beta^2 y^2 v^2/x_{cl}^2$.
The integral is finite at $a = 0$,
reproducing the free energy of a gas of free fermions at high temperatures:
\beq
\label{eq:I0}
I(0; V) = 2\, {\pi^2\over 12}\: T^2 \, .
\eeq
This high temperature result is independent of the background field,
as expected.

The first derivative of eq.\ref{eq:I} is:
\beq
\label{eq:dI}
{\partial I\over\partial a^2} \, (\, a^2\, ;\, V\, ) \: =\:
-\, {\sqrt{-V}\over e^a + 1} \: - \:
{1\over 2}\, (1 + V)\:\int_0^{\infty}
{d\, x^2 \over \sqrt{ (x^2 - a^2 V)\, (x^2 + a^2) }}\,
{1\over e^{\sqrt{x^2 + a^2}} + 1}
\eeq
after an integration by parts.
The surface term is $O(a^0)$ but the integral is
logarithmically divergent as $a^2\rightarrow 0$, so one can expect:
\beq
\label{eq:log}
{\partial I\over\partial a^2} \, (\, a^2\, ;\, V\, ) \: \sim \:
-{1\over 2}\,\sqrt{-V}\: +\: const.\,\log{a}
\eeq
near $a^2 = 0$.

To determine this non-analytic term,
the integral is broken into a series by use of the identity:
\beq
\label{eq:identity}
{1\over e^y + 1} \: = \:
{1\over 2} \: - \: \sum_{-\infty}^{+\infty}
{y\over y^2 + \pi^2 (2n + 1)^2}
\eeq
which follows from a contour integration of the left hand side\cite{DOL}.
Now, the integrals over each term in this expansion are divergent,
so they will be regulated by introducing:
\beq
\label{eq:J}
J_{\epsilon}\, (a^2; V)  \, \equiv\,
\int_0^{\infty}
{d\, x^2 \over \sqrt{ (x^2 - a^2 V)\, (x^2 + a^2) }}
{x^{-\epsilon}\over e^{\sqrt{x^2 + a^2}} + 1}
\eeq
and
\beq
\label{eq:J1J2}
J_{\epsilon}\, (a^2; V)  \, \equiv\,
J^{(1)}_{\epsilon}\, (a^2; V) \: + \:  J^{(2)}_{\epsilon}\, (a^2; V) .
\eeq
where
\beq
\label{eq:J1}
J^{(1)}_{\epsilon}\, (a^2; V)  \,\equiv\,
-\,\sum_{-\infty}^{+\infty}\:
\int_0^{\infty}
d\, x^2 \, {x^{-\epsilon}\over \sqrt{ x^2 - a^2 V}}
{1\over x^2 + a^2 + \pi^2 (2n + 1)^2}
\eeq
and
\beq
\label{eq:J2}
J^{(2)}_{\epsilon}\, (a^2; V)  \,\equiv\,
{1\over 2}\, \int_0^{\infty}
d\, x^2 \:
{x^{-\epsilon}\over \sqrt{ (x^2 - a^2 V)\, (x^2 + a^2) }} \, .
\eeq

A little analysis shows that $J^{(1)}$ and $J^{(2)}$ converge if
$0 < \epsilon < 2$.
The integrals will be estimated for small $a^2$ in this region,
and then analytically continued to $\epsilon = 0$,
which is a regular point of the original integral, eq.\ref{eq:J}.
The result is the unique high-T expansion of $I$.

First, eq.\ref{eq:J1} can be rewritten as:
\beqa
\label{eq:J1rewrite}
J^{(1)}_{\epsilon}\, (a^2; V)  & = &
-\,\sum_{-\infty}^{+\infty}\,
{1\over \left[ \pi^2 (2n + 1)^2\right]^{1 + \epsilon \over 2} } \,
{1\over \left[1 \, +\,  a^2/\pi^2 (2n + 1)^2\right]^{1 + \epsilon \over 2}} \,
\times \nonumber \\
&    &
\hspace{4cm} \int_0^{\infty}\, d\, u^2 \,
{ u^{-\epsilon}\over {(u^2 + 1) \sqrt{u^2 + c^2(n)}} }
\eeqa
where $c^2(n)$ is an $n$-dependent constant of order $a^2$.
To leading order in $a^2$ then, eq.\ref{eq:J1rewrite} is:
\beq
\label{eq:J1approximation}
J^{(1)}_{\epsilon}\, (a^2; V) \, = \,
-\,\sum_{-\infty}^{+\infty}\:
{1\over \left[ \pi^2 (2n + 1)^2\right]^{1 + \epsilon \over 2}} \:
\int_0^{\infty}\, d\, u^2 \:
{u^{-\epsilon}\over u(u^2 + 1)}
\:\left[ 1 \, + \, O(a^2) \right] \, .
\eeq
The remaining integral is elementary, and the sum can be expressed
in terms of the Riemann-Zeta function.
\beq
\label{eq:J1done}
J^{(1)}_{\epsilon}\, (a^2; V) =
-\, 2\,
\left( 1 \, - \, 1/2^{1 + \epsilon} \right)
\zeta \left[ 1 + \epsilon \right] \,
{\pi^{-\epsilon}\over \cos ({\pi\epsilon\over 2}) }
\: + \: O(a^2) \, .
\eeq
Finally, this result is continued to $\epsilon = 0$ by means
of the formula
\beq
\label{eq:zetacontinued}
\zeta \left[ 1 + \epsilon \right] \: = \:
-\,
{ 2^\epsilon \, \pi^{1 + \epsilon}\, \zeta \left[ -\epsilon \right]
\over \sin ({\pi\epsilon\over 2})\, \Gamma\left[1 + \epsilon\right] }
\eeq
which gives
\beq
\label{eq:J1expanded}
J^{(1)}_{\epsilon}\, (a^2; V)  \: =\:
-\, 1/\epsilon \, - \, \gamma \, + \, \log (\pi/2) \: +\:  O(\epsilon)
\: +\:  O(a^2) \, ,
\eeq
where $\gamma\simeq 0.53$ is the Euler Number.

Next, eq.\ref{eq:J2} can be rewritten as:
\beq
\label{eq:J2rewrite}
J^{(2)}_{\epsilon}\, (a^2; V) \, = \,
a^{-\,\epsilon}\: J^{(2)}_{\epsilon}\, (1; V)  \, = \,
\left(\, 1 \, - \, \epsilon\log a  \, + \, O(\epsilon^2) \,\right)
J^{(2)}_{\epsilon}\, (1; V) \, .
\eeq
This is the source of the non-analyticity near $a = 0$.

Now, $ J^{(2)}_{\epsilon}\, (1; V) $ is easily analyzed by
expressing  it in terms of two special functions\cite{GRA}.
\beq
\label{eq:J2special}
J^{(2)}_{\epsilon}\, (1; V) \, = \,
{1\over 2\sqrt{- V}}\:
B (1 - \epsilon/2, \epsilon/2) \:
_2 F_1 \left[ 1/2 , 1 - \epsilon/2 ; 1; 1 + 1/V \right] \, .
\eeq
The pole at $\epsilon = 0$ is contained in the Beta function,
$ B (1 - \epsilon/2, \epsilon/2) $.
The hypergeometric function,
$ _2 F_1  $, is analytic in its second argument near $1$.
Expanding in a Taylor series there gives
\beq
\label{eq:J2Taylor}
J^{(2)}_{\epsilon}\, (1; V) \, = \,
{1\over 2}\:
B (1 - \epsilon/2, \epsilon/2) \:
\left[\,
 1 \: -\: {\epsilon\over 2\sqrt{- V}}
\; _2 F_1' \left[ 1/2 , 1; 1; 1 + 1/V \right] \: + \:
O(\epsilon^2)
\,\right]
\eeq
where the prime denotes differentiation with respect to the
second argument of $_2 F_1$.
This function can be determined by using the
various recursion relations which relate
hypergeometric functions of different arguments.
The result is:
\beq
\label{eq:hyper}
_2 F_1' \left[ 1/2 , 1; 1; 1 + 1/V \right] \: = \:
2\,\sqrt{-V}\,\log \left[{1\over 2}\,\left( 1 \, +\, \sqrt{-V}\right)\right]
\, .
\eeq
Now,
expanding the Beta function for small $\epsilon$ in eq.\ref{eq:J2special}
yields:
\beq
\label{eq:J2done}
J^{(2)}_{\epsilon}\, (1; V) \: = \: 1/\epsilon \, - \,
\log \left[{1\over 2}\,\left( 1 \, +\, \sqrt{-V}\right)\right]
\, + \,O(\epsilon) \, .
\eeq
Finally,
eq.\ref{eq:J2rewrite} becomes
\beq
\label{eq:J2expanded}
J^{(2)}_{\epsilon}\, (a^2; V) \: = \: 1/\epsilon \, - \, \log (a)
\, - \,
\log \left[{1\over 2}\,\left( 1 \, +\, \sqrt{-V}\right)\right]
\, + \, O(\epsilon) \, .
\eeq

Combining eq.\ref{eq:J1expanded}, eq.\ref{eq:J2expanded} and
eq.\ref{eq:J1J2}, the poles in $\epsilon $ cancel and give
\beq
\label{eq:Jdone}
J_\epsilon\, (a^2; V) \: =\:
-\, \log (2 a/\pi)
\, -\, \gamma_E
\, -\, \log \left[{1\over 2}\,\left( 1 \, +\, \sqrt{-V}\right)\right]
\, +\, O(a)\, + \, O(\epsilon) \, .
\eeq
\beqa
\label{eq:dIdone}
{\partial I\over\partial a^2} \, (\, a^2\, ;\, V\, ) & = &
-\,{1\over 2}\,\sqrt{-V} \: + \:
{1\over 2}\,\left(1 + V\right)\,
\left\{ \: \rule{0mm}{5mm}
\log (2 a/\pi)
\, +\, \gamma_E  \right. \nonumber \\
&   & \hspace{3cm} \left.
+\: \log \left[{1\over 2}\,\left( 1 \, +\, \sqrt{-V}\right)\right]
\: \right\}
\, +\, O(a)\, + \, O(\epsilon) \, .
\eeqa
Now,
$I(a^2)$ can be recovered by
expanding the surface term in eq.\ref{eq:dI}
to lowest order in $a$,
combining with eq.\ref{eq:dIdone}, and
integrating with respect to
$a^2$.
\beqa
\label{eq:appresult}
I\, (\, a^2\, ;\, V\, ) \: =\:
I\, (\, 0\, ;\, V\, )
& - &
{1\over 2}\, a^2\,\sqrt{-V} \, + \,
{1\over 4}\,\left(1 + V\right)\, a^2\left\{\,
\log (a^2/\pi^2) + 2 \gamma_E - 1
\,\right\} \nonumber \\
& + &
{1\over 2}\,\left(1 + V\right)\, a^2\,
\log \left[\, 1 \, +\, \sqrt{-V}\,\right]
\, + \, O(a^4)
\eeqa
where the integration constant is determined by eq.\ref{eq:I0}.
This result establishes the free energy in eq.\ref{eq:Fintapprox}.

\section{Acknowledgements}

The authors gratefully acknowledge
the support of the Aspen Center for Physics,
where this work was initiated.
The authors also thank S.D.H. Hsu and S.G. Naculich
for helpful conversations.

\vskip.3in
\centerline{{\bf Figure Captions}}

\noindent Figure 1: The ratio of the size of the thermal sphaleron to the zero
temperature sphaleron as a function of the parameter $a=\beta y/x_{cl}$.

\noindent Figure 2: The free enrgy of the thermal sphaleron as a function
of the parameter $a=\beta y/x_{cl}$.

\end{document}